\documentclass[
    ,final            
  ]
  {aipproc}

\layoutstyle{6x9}
\newcommand{\be}{\begin{equation}}
\newcommand{\ee}{\end{equation}}
\newcommand{\bea}{\begin{eqnarray}}
\newcommand{\eea}{\end{eqnarray}}

\newcommand{\D}{\displaystyle}
\newcommand{\g}{\gamma}
\newcommand{\f}{\frac}

\newcommand{\intc}[1]{{\int\frac{d#1}{2i\pi}}}
\newcommand\lr[1]{{\left({#1}\right)}}

\begin{document}

\title{Small-x effects in forward-jet production at HERA}


\author{Cyrille Marquet}{
  address={Service de physique th{\'e}orique, CEA/Saclay,
  91191 Gif-sur-Yvette cedex, France\\
  URA 2306, unit{\'e} de recherche associ{\'e}e au CNRS
  }}



\begin{abstract}
We investigate small$-x$ effects in forward-jet production at HERA in the 
two-hard-scale region $k_T\!\sim\!Q\!\gg\!\Lambda_{QCD}$. We show that, despite 
describing 
different energy regimes, both a BFKL parametrization and saturation 
parametrizations describe well the H1 and ZEUS data for $d\sigma/dx$ published a 
few years ago. This is confirmed when comparing the predictions to the latest 
data.
\end{abstract}

\maketitle


\section{Introduction}

Forward-jet production is a process in which a virtual photon strongly interacts 
with a proton and a jet is detected in the forward direction of the proton. The 
virtuality of the photon $Q^2$ and the squared transverse momentum of the jet 
$k_T^2$ are hard scales of about the same magnitude. In the Regge limit of 
perturbative QCD, {\em i.e.} when the centre-of-mass energy in a collision is 
much bigger than the fixed hard scales of the problem, the scattering amplitudes 
grow with increasing energy as described by the BFKL equation~\cite{bfkl}. The 
forward-jet measurement was originally proposed~\cite{dis} to test the BFKL 
equation because, if the energy in the photon-proton collision $W$ is large 
enough, it lies in the kinematic regime corresponding to the Regge limit 
($W^2\!\gg\!Q^2$). 

The question is whether the BFKL equation is relevant at the present energies, 
or if usual perturbative QCD in the Bjorken limit is still sufficient to 
describe the data. We adress that problem by computing the forward-jet 
cross-section in the high-energy regime and by comparing the BFKL predictions 
with the available data. We also adress the problem of saturation~\cite{glr}: it 
is well-known that the BFKL growth is damped by saturation effects when energies 
become too high and the scattering amplitudes approach the unitatity limit. We 
implement saturation effects in a very simple way, inspired by the Golec-Biernat 
and W\"usthoff approach~\cite{golec} and check the consistency with the data.

\section{Formulation}

The QCD cross-section for forward-jet production in a lepton-proton collision
reads
\be
\f{d^{(4)}\sigma}{dxdQ^2dx_Jdk_T^2}=\f{\alpha_{em}}{\pi xQ^2}
\left\{\lr{\f{d\sigma^{\g*p\!\rightarrow\!JX}_T}{dx_Jdk_T^2}
+\f{d\sigma^{\g*p\!\rightarrow\!JX}_L}{dx_Jdk_T^2}}(1-y)
+\f{d\sigma^{\g*p\!\rightarrow\!JX}_T}{dx_Jdk_T^2}\f{y^2}2\right\}\ 
,\label{one}\ee
where $x$ and $y$ are the usual kinematic variables of deep inelastic 
scattering and $Q^2$ is the virtuality of the intermediate photon that undergoes 
the hadronic interaction. 
$d\sigma^{\g*p\!\rightarrow\!JX}_{T,L}/dx_Jdk_T^2$ is 
the cross-section for forward-jet production in the collision of this 
transversally (T) or longitudinally (L) polarized virtual photon with the target 
proton. $k_T$ is the jet transverse momentum and $x_J$ its longitudinal momentum 
fraction with respect to the proton.

Let us now consider the high-energy regime: 
$x\!=\!\log(Q^2/(Q^2\!+\!W^2))\!\ll\!1.$
In an appropriate frame called the dipole 
frame, the virtual photon undergoes the hadronic interaction via a fluctuation 
into a colorless $q\bar q$ pair, a dipole. The squared wavefunctions 
$\phi^\gamma_T$ and $\phi^\gamma_L$ describing the splitting of the virtual 
photon onto a dipole are well-known. The dipole then interacts with the target 
proton and one has the following factorization
\be
\f{d\sigma^{\g*p\!\rightarrow\!JX}_{T,L}}{dx_Jdk_T^2}=\int_0^\infty 2\pi rdr\  
\phi_{T,L}^{\g}(r,Q)
\f{d\sigma_{q\bar q}}{dx_Jdk_T^2}(r)\ .\label{fact}\ee
$d\sigma_{q\bar q}(r)/dx_Jdk_T^2$ is the cross-section for forward-jet 
production in the dipole-proton collision. The integration variable $r$ 
represents the size of the intermediate dipole.

It was shown in~\cite{marq} that the emission of the forward jet can be 
described through the interaction of an effective gluonic (gg) dipole:
\be
\f{d\sigma_{q\bar q}}{dx_Jdk_T^2}(r)=\f{\pi N_c}{16k_T^2}f_{eff}(x_J,k_T^2)
\int_0^\infty d\bar{r}\ J_0(k_T\bar{r})\ \f{\partial}{\partial 
\bar{r}}\lr{\bar{r}
\f{\partial}{\partial \bar{r}}\ \sigma_{(q\bar q)(gg)}(r,\bar{r},Y)}
\label{fjdcs}\ee
with $Y\!=\!\log(x_J\!/\!x)$ the rapidity assumed very large. 
$\sigma_{(q\bar q)(gg)}(r,\bar{r},Y)$ is the $q\bar q$ dipole (size $r$)-$gg$ 
dipole (size $\bar{r}$) total cross-section with rapidity Y. As usual, the 
dipoles emerge as the effective degrees of freedom at high energies:
$\sigma_{(q\bar q)(gg)}$ contains any number of gluon exchanges and therefore 
this formulation goes beyond $k_T-$factorization which assumes only a two-gluon 
exchange. The effective parton distribution function $f_{eff}$ is given by: 
$f_{eff}(x_J,k_T^2)=g(x_J,k_T^2)\!+\!C_F\lr{q(x_J,k_T^2)\!+\!\bar{q}(x_J,k_T^2)}
/N_c$ where $g$ (resp. $q$, $\bar{q}$) is the gluon (resp. quark, antiquark) 
distribution function in the incident proton.

\subsection{BFKL parametrization}

The BFKL $q\bar q-$dipole $gg-$dipole cross-section reads
\be
\sigma^{BFKL}_{(q\bar q)(gg)}(r,\bar{r},Y)=2\pi\alpha_s^2r^2
\intc{\g}\lr{\f{\bar{r}}r}^{2\g}\
\f{\exp{\lr{\D\f{\alpha_sN_c}{\pi}\chi(\g)Y}}}{\g^2(1\!-\!\g)^2}
\label{sdd}
\ee
with the complex integral running along the 
imaginary axis from $1/2\!-\!i\infty$ to $1/2\!+\!i\infty$ and
with the BFKL kernel given by 
$\chi(\g)\!=\!2\psi(1)\!-\!\psi(1\!-\!\g)\!-\!\psi(\g)$
where $\psi(\g)$ is the logarithmic derivative of the Gamma function.
It comes about when the interaction between the $q\bar q-$dipole and the 
$gg-$dipole is restricted to a two-gluon exchange. One can easily show, putting 
(\ref{sdd}) in (\ref{fact}) and (\ref{fjdcs}), that this formulation is 
equivalent to using $k_T-$factorization. We are going to perform a fit of the 
parametrization (\ref{sdd}) to the data. The parameters are 
$\lambda\!=\!4\alpha_sN_c\log(2)/\pi$ and a normalization.

\begin{figure}[htb]
\begin{minipage}[t]{70mm}
\includegraphics[width=6.5cm,clip=true]{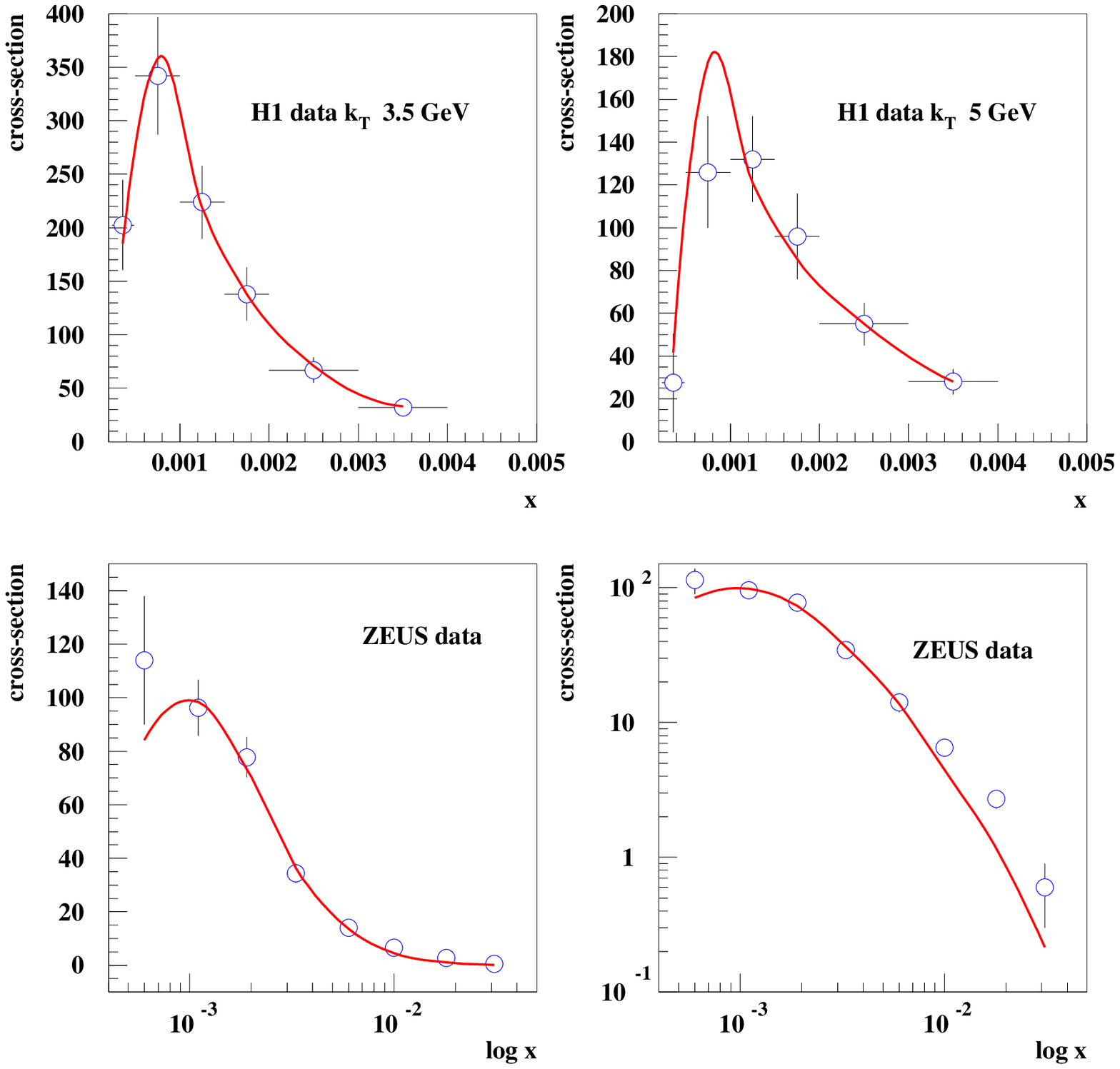}
\end{minipage}
\hspace{\fill}
\begin{minipage}[t]{70mm}
\includegraphics[width=6.5cm,clip=true]{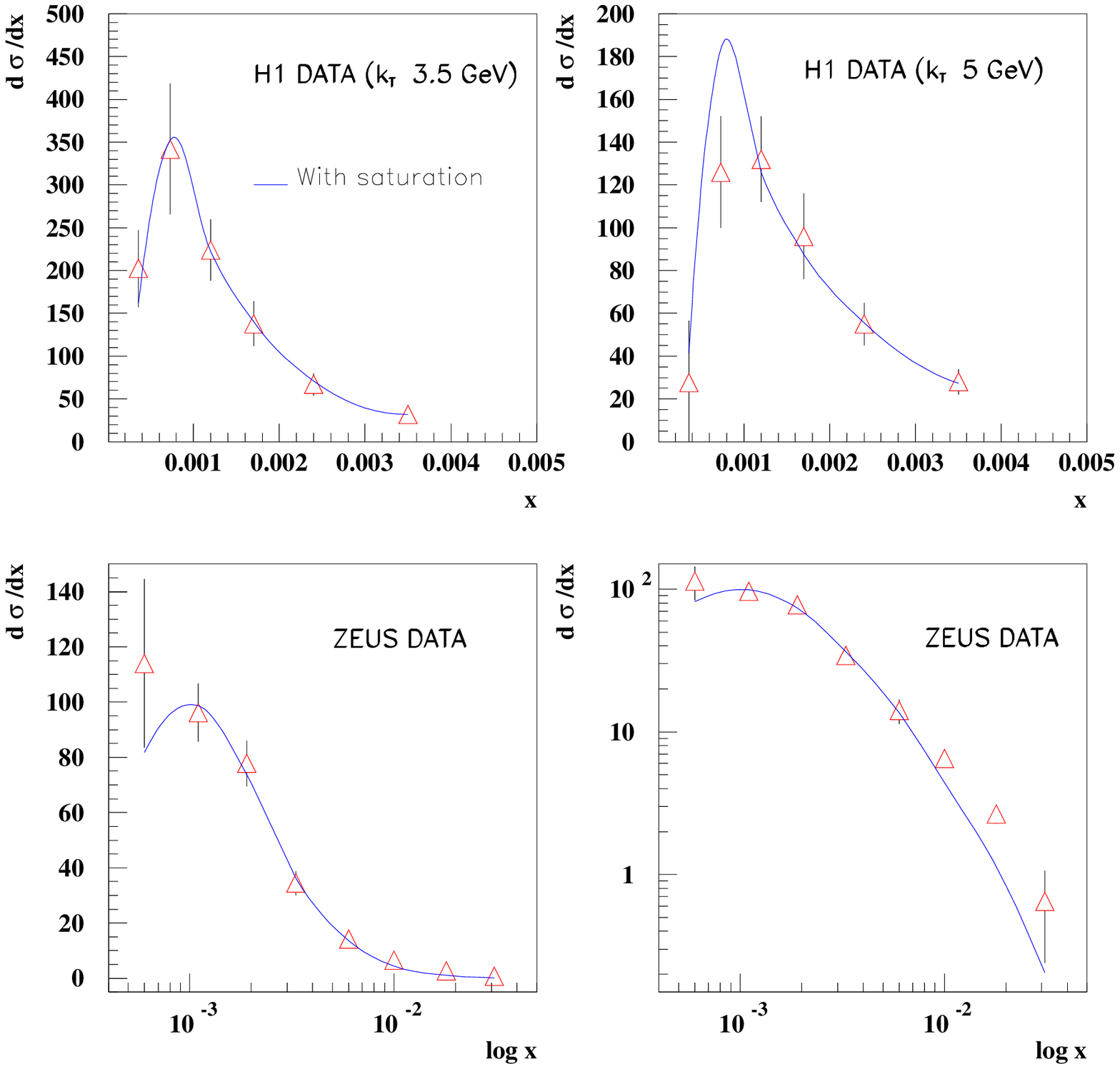}
\end{minipage}
\caption{Fits to the H1 and ZEUS forward-jet old data for $d\sigma/dx.$ The left 
plot shows the BFKL fit and the right plot shows one of the saturation fits 
(called sat. in the text).}
\label{fits}
\end{figure}

\subsection{Saturation parametrization}

To take into account saturation effects, let us consider the following 
parametrization:
\be
\sigma^{sat}_{(q\bar q)(gg)}(r,\bar{r},Y)=4\pi\alpha_s^2\sigma_0
\lr{1-\exp\lr{-\f{r_{\rm eff}^2(r,\bar{r})}{4R_0^2(Y)}}}\ .
\label{sigmadd}
\ee
The dipole-dipole {\it effective} radius $r^2_{\rm  eff}(r,\bar{r})$ is defined 
through the two-gluon exchange:
\be 
4\pi\alpha_s^2r^2_{\rm eff}(r,\bar{r})\equiv
\sigma^{BFKL}_{(q\bar q)(gg)}(r,\bar{r},0)
=4\pi\alpha_s^2\min(r^2\!,\bar{r}^2)\left\{1\!+\!\log
\frac{\max(r\!,\bar{r})}{\min(r\!,\bar{r})}\right\}\label{reff}\ee
while the saturation radius is parametrized by 
$R_0(Y)\!=\!e^{-\f{\lambda}2\left(Y-Y_0\right)}/Q_0$ with $Q_0\!\equiv\!1\ GeV.$ 
The parameters for the fit are $\lambda,$ $Y_0$ and the normalization 
$\sigma_0$.

\section{Phenomenology}

\begin{figure}[htb]
\begin{minipage}[t]{70mm}
\includegraphics[width=6.5cm,clip=true]{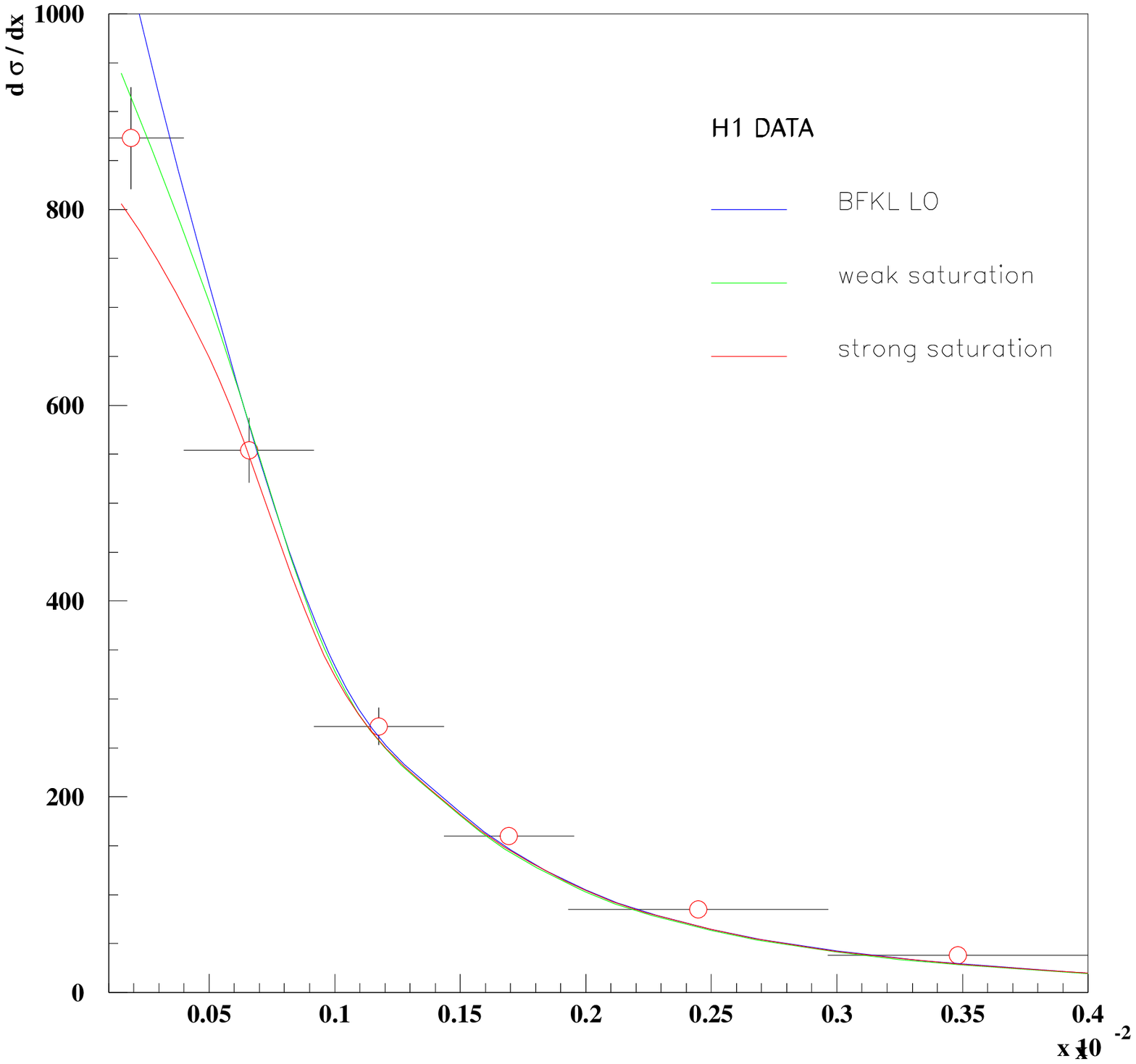}
\end{minipage}
\hspace{\fill}
\begin{minipage}[t]{70mm}
\includegraphics[width=6.5cm,clip=true]{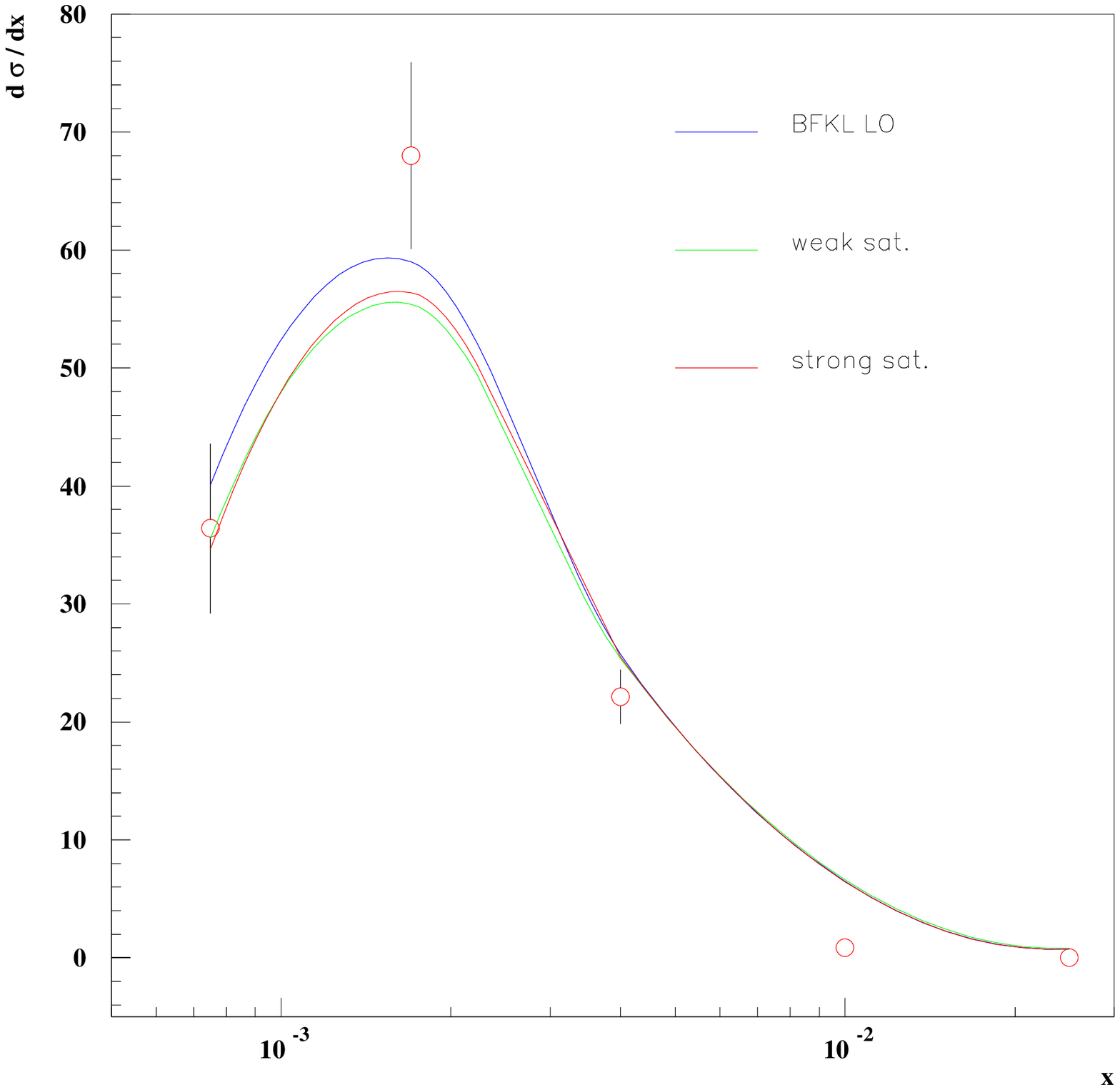}
\end{minipage}
\caption{Comparisons between the H1 (left plot) and ZEUS (right plot) 
forward-jet new data for $d\sigma/dx$ and the BFKL and saturation 
parametrizations.}
\label{comp}\end{figure}

To compare the cross-section (\ref{one}) with the data for $d\sigma/dx,$ the 
three remaining integration are carried out taking into account the different 
sets of cuts provided by the different experiments. Fits have been performed to 
the old sets of data~\cite{h1,zeus} for the BFKL parametrization~\cite{fjets} 
and the saturation parametrization~\cite{mpr}. With all $\chi^2$ values of about 
1, the BFKL fit gives $\lambda\!=\!0.430$ and the saturation fit shows two 
$\chi^2$ minima for $(\lambda\!=\!0.402,Y_0\!=\!-0.82)$ (sat.) and 
$(\lambda\!=\!0.370,Y_0\!=\!8.23)$ (weak sat.). The plots are shown on Fig1. 
Despite describing 
different energy regimes, both a BFKL parametrization and saturation 
parametrizations describe well the data. The first saturation minima corresponds 
to a strong saturation effect as, for typical values of $Y,$ the saturation 
scale $1/R_0$ is 5 Gev which is the value of a typical $k_T.$ The second 
saturation minima corresponds to small saturation effets and rather describes 
BFKL physics.

Let us now look at the new data~\cite{h1new,zeusnew} for $d\sigma/dx$ which go 
to lower $x$. Without performing any new fit of the parameters, but rather by 
taking the values already obtained, the three parametrizations describe very 
well the new data, as shown on Fig2. One cannot really distinguish between the 
three curves even if at small values of $x$, one starts to see the difference 
between them. At the lowest values of $x,$ NLOQCD predictions are about a factor 
1.5 to 2.5 below the data depending on the experiment and the error bars. 
However, it could be that adding a resolved-photon component to the NLO 
predictions will pull them within the uncertainties which moderates the 
conclusion that the BFKL resummation is needed to describe those data. The fact 
that two saturation parametrizations are consistent with the data also asks for 
further study. 

We intend to complete our analysis~\cite{inprep} by considering the other 
measurements $d\sigma/dQ^2$ and $d\sigma/dk_T$ by ZEUS and 
$d\sigma/dxdQ^2dk_T^2,$ by H1 and Mueller-Navelet jets~\cite{mnj} at Tevatron or 
LHC. These could help clarifing the situation~\cite{marpes}. 


 



\end{document}